\documentclass[12pt,a4paper]{article}

\usepackage{latexsym}
\usepackage{amsmath}
\usepackage{amsfonts}
\usepackage{amssymb}

\def\G{{\cal G}}                        %
\def\M{{\cal M}}                        %
\def\D{{\cal D}}                        %
\def\L{{\cal L}}                        %
\def\A{{\cal A}}                        %
\def\C{{\cal C}}                        %
\def\K{{\cal K}}                        %
\def\E{{\cal E}}                        %
\def\B{{\cal B}}                        %
\def\cL{{\check{\cal L}^*}}             %
\def\cK{{\check{\cal K}^*}}             %
\def\ad{{\mathrm{ad}\,}}                %
\def\bZ{{\mathbb Z}}                    %
\def\bC{{\mathbb C}}                    %
\def\bR{{\mathbb R}}                    %
\def\bN{{\mathbb N}}                    %

\begin{document}

\begin{titlepage} 
\vspace*{0.5cm}
\begin{center}

{\large\bf  On dynamical $r$-matrices obtained from Dirac reduction \\

and their generalizations to affine Lie algebras}

\end{center}

\vspace{1.5cm}

\begin{center}
{\large L. Feh\'er\footnote{Corresponding author, E-mail: lfeher@sol.cc.u-szeged.hu}, 
A. G\'abor and B.G. Pusztai}
\end{center}
\bigskip
\begin{center}
Department of Theoretical Physics,  University of Szeged \\
Tisza Lajos krt 84-86, H-6720 Szeged, Hungary \\
\end{center}

\vspace{2.2cm}

\begin{abstract}  

According to Etingof and Varchenko, 
the classical dynamical Yang-Baxter equation is a guarantee for  
the consistency of the Poisson bracket on certain Poisson-Lie groupoids.
Here it is noticed that Dirac reductions of these Poisson manifolds 
give rise to a mapping from  dynamical 
$r$-matrices on a pair $\L\subset \A$ to those on another pair
$\K\subset \A$, where $\K\subset \L\subset \A$ is a chain
of Lie algebras for which $\L$ admits a reductive decomposition as
$\L=\K+\M$.
Several known dynamical $r$-matrices appear naturally in this setting, 
and its application provides new $r$-matrices, too. 
In particular, we exhibit a family of $r$-matrices for which 
the dynamical variable lies in the grade zero subalgebra of 
an extended affine Lie algebra obtained from a twisted
loop algebra based on an arbitrary finite dimensional self-dual Lie algebra.

\end{abstract}

\end{titlepage}

\section{Introduction}
\setcounter{equation}{0}

The Yang-Baxter equation and the associated algebraic structures 
play a central role in the theory of integrable systems.
Recently there has been growing interest in dynamical generalizations 
of these objects (for a review, see \cite{ES}).
Our concern in this paper is the classical dynamical 
Yang-Baxter equation (CDYBE) that originally appeared in studies 
of the Liouville-Toda and the WZNW conformal field theories \cite{Gervais,BDF,Feld}.
In its general form the CDYBE is defined \cite{EV} as follows.
Let ${\cal A}$ be a Lie algebra and ${\cal L}\subset {\cal A}$ a Lie 
subalgebra with dual space ${\cal L}^*$.
A dynamical $r$-matrix with respect to the pair ${\cal L}\subset {\cal A}$
is a map $r$ from an open domain $\cL \subset {\cal L}^*$
to ${\cal A}\otimes {\cal A}$ subject to 
\begin{equation}
[r_{12}, r_{13}] + 
L^1_a \frac{\partial r_{23}}{\partial \lambda_a} +
\hbox{cycl. perm.} =0.
\label{1.1}\end{equation}
Here the $\lambda_a$ are coordinates on $\L^*$ with respect to 
a basis $\{ L_a\}$ of ${\cal L}$, 
the usual tensorial notation and the summation convention are used
throughout the paper.
The cyclic permutations act on the three tensorial factors,
for any $r=X^i \otimes Y_i \in \A\otimes \A$ one defines 
$r_{12} = X^i \otimes Y_i \otimes 1$, 
$r_{31}=Y_i \otimes 1 \otimes X^i$ and so on.
It is further required that the symmetric part of $r$ 
is an  $\A$-invariant constant
element of ${\cal A}\otimes {\cal A}$ 
and the function $r: \cL\rightarrow {\cal A}\otimes {\cal A}$
is equivariant with respect to the natural infinitesimal actions of
${\cal L}$ on the respective spaces.
Etingof and Varchenko \cite{EV} found an interesting geometric interpretation
of the CDYBE that generalizes Drinfeld's interpretation of the CYBE 
in terms of Poisson-Lie groups \cite{Drinfeld}.
Namely, they constructed a so called dynamical Poisson-Lie groupoid 
structure on the direct product manifold
\begin{equation}
\check \L^* \times A \times \check \L^* 
\label{1.2}\end{equation}
where $A$ is a connected Lie group with Lie algebra $\A$.
The Poisson structure on (\ref{1.2}) is encoded by $r$  
in such a way that the antisymmetry and the Jacobi identity 
enforce the above 
mentioned invariance and equivariance properties of $r$ 
together with the condition that the function 
\begin{equation}
CDYB(r):= 
[r_{12}, r_{13}] + L^1_a \frac{\partial r_{23}}{\partial \lambda_a} +\hbox{cycl. perm.}
\label{1.3}\end{equation}
must yield an $A$-invariant constant element of 
$\A\wedge \A \wedge \A$.
Many examples and a classification 
of the meromorphic classical dynamical $r$-matrices 
for certain choices of the pair
${\cal L}\subset {\cal A}$  are now available \cite{ES,EV}.

The purpose of this paper is to point out a simple mechanism 
whereby some known and some new solutions of the CDYBE 
can viewed from a unified perspective.
Our basic idea is that the imposition of suitable constraints 
on the dynamical Poisson-Lie groupoid (\ref{1.2}) 
will result in a reduced Poisson-Lie groupoid of the form
 \begin{equation}
\cK \times A \times \cK
\label{1.4}\end{equation}
for some subalgebras $\K\subset \L$.
The Dirac bracket defined by the reduction
will be encoded by an $r$-matrix $r^*: \check \K^* \rightarrow \A\otimes \A$
that solves the CDYBE for the pair $\K\subset \A$ whenever the original $r$-matrix 
solves it for the pair  $\L\subset \A$.
It will be shown that the reduction works in this manner
if $\K\subset \L$ admits a $\K$ invariant complementary 
linear space  
and the constraints of the reduction are second class.
Under these conditions, we obtain a simple formula for $r^*$ by applying   
the standard formula to determine the Dirac bracket on
$\check \K^* \times A \times \check \K^*$.  
This formula implies that
$CDYB(r)=CDYB(r^*)$, and therefore the reduction 
closes on classical dynamical $r$-matrices.

Our remark on the Dirac reduction of dynamical $r$-matrices 
complements the known constructions of solutions of the CDYBE
and sheds a new light on the origin of some solutions.
For instance, if the pair ${\cal L}\subset {\cal A}$ 
is given by the Cartan subalgebra of a simple Lie algebra,
which is a case of principal interest,  
then the corresponding basic 
rational and trigonometric solutions can be viewed  
as Dirac reductions of respectively the zero 
and the  so called  `canonical' (or Alekseev-Meinrenken) $r$-matrices \cite{EV, AM, BFP}
for which $\L=\A$.
We note that 
an equivalent result can be extracted from \cite{EV} as well
(see Theorem 3.14 in \cite{EV}).
However,  Dirac reduction is not mentioned in \cite{EV},
and it works in more general circumstances 
than those considered in this reference. 
In particular, in equation (\ref{4.7}) a class of $r$-matrices is displayed  
which is applicable  to arbitrary (not necessarily simple or reductive) 
finite dimensional self-dual Lie algebras \cite{Figu}. 
To illustrate that formula (\ref{4.7}) contains new  
dynamical $r$-matrices, too, we shall  apply it 
to the self-dual extension \cite{Sfetsos} of the Euclidean Lie
algebra $e(d)$ for even $d$. 
Moreover, we shall show that this formula remains well-defined in certain 
infinite dimensional situations as well.
In fact, 
several new $r$-matrices will be obtained by applying (\ref{4.7}) in the cases 
for which the dynamical variable lies in the grade zero
subalgebra of an extended affine Lie algebra associated with a twisted
loop algebra based on an arbitrary finite dimensional self-dual Lie algebra. 
These yield generalizations of Felder's spectral parameter 
dependent dynamical $r$-matrices \cite{Feld}
upon applying evaluation homomorphisms to the twisted loop algebras.

The organization of the paper is the following.
A short recall  of the geometric 
interpretation of the CDYBE from \cite{EV} is presented in Section 2.
Section 3 is devoted to the Dirac reduction of dynamical $r$-matrices.
In Section 4 examples are given on arbitrary finite dimensional self-dual
Lie algebras, and some of these $r$-matrices are
generalized to affine Lie algebras in Section 5. 
The final section contains a discussion of the results, open questions
and comments on the literature.

The main results are given by Proposition 1 in Section 3,
formula (\ref{4.7}) in Section 4, and Proposition 2 in Section 5.
We consider the development of the Dirac reduction viewpoint to be our
most important result, since it may lead to further results in the future.
For example, it should be possible to apply Hamiltonian reduction {\em after} 
quantizing  the Poisson-Lie groupoids that underlie 
the dynamical $r$-matrices,
since the second class constraints that appear in  the examples
usually admit a natural separation into first class constraints and 
gauge fixing conditions.

\section{Geometric interpretation of the CDYBE}
\setcounter{equation}{0}

We wish to apply Dirac reduction to the dynamical 
Poisson-Lie groupoids that encode the dynamical $r$-matrices.
As a preparation, we here recall from \cite{EV} the definition of these
Poisson manifolds in a form convenient for our purpose.

Let us denote the elements of the space in (\ref{1.2}) as 
\begin{equation}
\cL \times A \times \cL=\{ (\lambda^F, g, \lambda^I)\},
\label{2.1}\end{equation}
and let $\lambda_a:= \lambda(L_a)$ be the components of $\lambda\in \cL$
with respect to 
a basis $L_a$ of $\L$ for which 
\begin{equation}
[L_a, L_b]= f_{ab}^{\phantom{ab}c} L_c.
\label{2.2}\end{equation}
Consider a function
$r: \cL \rightarrow \A\otimes \A$,
and equip  
$\cL \times A \times \cL$ with a Poisson bracket $\{\ ,\ \}$ 
of the following form:
\begin{eqnarray}
&&
\{ g_1, g_2\} = g_1 g_2 r(\lambda^I) - 
r(\lambda^F) g_1 g_2 
\nonumber\\
&& \{ g, \lambda^I_a\} = g L_a
\nonumber\\
&& \{ g, \lambda^F_a\} = L_a g
\nonumber\\
&& \{ \lambda^I_a, \lambda^I_b\} =-f_{ab}^{\phantom{ab}c}
\lambda^I_c
\nonumber\\
&& \{ \lambda^F_a, \lambda^F_b\} =f_{ab}^{\phantom{ab}c}
\lambda^F_c
\nonumber\\
&&\{ \lambda^I_a, \lambda^F_b\} =0.
\label{2.3}\end{eqnarray}
In this formula $g_1 := g\otimes 1$ and $g_2:= 1\otimes g$ 
are really defined  in terms of matrix representations of the group $A$.
If one fixes a representation, then the first line of (\ref{2.3}) serves 
to define the value of
$\{ g_1, g_2\}_{ij,kl}= \{g_{ij}, g_{kl}\}$, while the second line 
means that $\{ g_{ij}, \lambda^I_a\} = (g L_a)_{ij}$. 
The antisymmetry and the Jacobi identity of the Poisson bracket
lead to the requirements on $r$ mentioned in the introduction
as follows \cite{EV}.
First, the antisymmetry $\{ g_1, g_2\} = - \{ g_2, g_1\}$ 
requires 
\begin{equation}
r^s := \frac{1}{2} (r + r_{21}) 
\label{2.4}\end{equation}
to be an $A$-invariant constant element of $\A\otimes \A$.
Second, the Jacobi identities
\begin{equation}
\{\{ g_1, g_2\}, \lambda_a^I\} + \hbox{cycl. perm.} =0=
\{ \{ g_1, g_2\}, \lambda_a^F\} + \hbox{cycl. perm.} 
\label{2.5}\end{equation}
are equivalent to the condition
\begin{equation}
[ L_a^1 + L_a^2, r(\lambda) ]= f_{ba}^{\phantom{ba}c}\lambda_c 
\frac{\partial r(\lambda)}{ \partial \lambda_b},
\label{2.6}\end{equation}
which is the coordinatewise description 
of the $\L$-equivariance of the map $r$.
This equation further restricts only the antisymmetric part $r^a$ of $r=(r^s + r^a)$.
Third, an easy calculation gives 
\begin{equation}
\{\{ g_1, g_2\}, g_3\} + \hbox{cycl. perm.}
= \left(CDYB(r)(\lambda^F)\right) G -  G \left(CDYB(r)(\lambda^I)\right)
\label{2.7}\end{equation}
with $G:= g\otimes g\otimes g$.
This means that $CDYB(r)$ must be an $A$-invariant constant element of 
$\A\otimes \A\otimes \A$.

We saw that (\ref{2.3}) is indeed a Poisson bracket if and only if  $r^s$ and $CDYB(r)$
are $A$-invariant constants and (\ref{2.6}) holds. 
If $r^s$ is an $A$-invariant constant, then 
\begin{equation}
CDYB(r^a + r^s) = CDYB(r^a) + CDYB(r^s)= CDYB(r^a) + [r^s_{12}, r^s_{13}].
\label{2.8}\end{equation}
One sees from this that $CDYB(r)$ belongs to
$\A\wedge\A\wedge \A\subset \A\otimes \A\otimes\A$. 
Clearly, $r^s$ drops out from the Poisson bracket (\ref{2.3}). 
Its sole role is that in many cases one can achieve 
$CDYB(r)=0$ by adding a suitable $r^s$ to an $r^a$ for which
$CDYB(r^a)$ is a nonzero constant. 

Below we use only the above mentioned features 
of the Poisson manifold (\ref{2.1}). The form of the Poisson 
bracket (\ref{2.3}) guarantees that (\ref{2.1}) is a Poisson-Lie groupoid 
in the sense of Weinstein \cite{Wei}.
This is readily verified from the definitions, but   
is not directly relevant for the purposes of this paper (see \cite{EV}). 
Note  that the Poisson bracket (\ref{2.3}) 
is also valid in the trivial case for which $r=0$,
and we shall see that the Dirac reduction of this case 
leads to dynamical r-matrices for which $CDYB(r^a)=0$.

\section{Dirac reduction acting on dynamical $r$-matrices}
\setcounter{equation}{0}

We wish to reduce the phase space (\ref{2.1}), (\ref{2.3}) to an object 
of the similar kind (\ref{1.4}) with respect to a subalgebra 
$\K\subset \L$.
 For the reduction to work, we need two assumptions.
The first assumption is that $\K$ admits an invariant complementary
linear space $\M$ in $\L$, that is we have
\begin{equation} 
\L= \K + \M, \qquad [\K, \M] \subset \M. 
\label{3.1}\end{equation}
In this case we can choose an adapted basis of $\L$ as
\begin{equation}
\{ L_a\} = \{ K_i\} \cup \{ M_{\alpha}\}, 
\qquad 
K_i \in \K,
\quad
M_\alpha\in \M.
\label{3.2}\end{equation}
Correspondingly, the structure constants of $\L$ become
\begin{equation}
[K_i, K_j]= f_{ij}^{\phantom{ij}k} K_k,
\quad
[K_i, M_\alpha]= f_{i\alpha}^{\phantom{i\alpha}\beta} M_\beta,
\quad
[M_\alpha, M_\beta] =  f_{\alpha\beta}^{\phantom{\alpha\beta}\gamma} M_\gamma
+ f_{\alpha\beta}^{\phantom{\alpha\beta}i} K_i.
\label{3.3}\end{equation}
We also have the induced decomposition
\begin{equation}
\L^* = \K^* + \M^*, 
\qquad 
\K^*:= \M^\perp,
\quad
\M^*:= \K^\perp.
\label{3.4}\end{equation} 
Accordingly, we decompose any $\lambda \in \L^*$ as
\begin{equation}
\lambda= \kappa + \mu
\quad\hbox{with}\quad 
\kappa \in \K^*,
\quad
\mu\in \M^*,
\label{3.5}\end{equation}
and these constituents  have the components
\begin{equation}
\kappa_i = \kappa(K_i) = \lambda(K_i)= \lambda_i 
\quad\hbox{and}\quad
\mu_\alpha = \mu(M_\alpha) = \lambda(M_\alpha)=\lambda_\alpha.
\label{3.6}\end{equation} 

We define the reduction by 
putting the $\M^*$-components of $\lambda^I$ and $\lambda^F$ to zero,
i.e., we impose the constraints
\begin{equation}
\lambda^I_\alpha=0 
\quad\hbox{and}\quad
\lambda^F_\alpha =0.
\label{3.7}\end{equation}
We want these constraints to be second class in the Dirac sense \cite{Dirac}.
Clearly, this means that the function 
\begin{equation}
\C_{\alpha\beta}(\kappa):= -
f_{\alpha\beta}^{\phantom{\alpha\beta}i} \kappa_i
\label{3.8}\end{equation}
must define an invertible matrix.
Our second assumption is that 
this condition holds after a possible restriction of the domain $\cL$.
More precisely, we assume that 
\begin{equation}
\cK:= \{ \kappa \in \cL\cap \K^*\,\vert\, \C(\kappa): \hbox{invertible}\,\}
\neq \emptyset ,
\label{3.9}\end{equation}  
i.e., that  $\cK$ is a {\em nonempty} open submanifold of $\K^*$. 
The inverse of the matrix $\C_{\alpha\beta}(\kappa)$ 
will be denoted by $\D^{\alpha\beta}(\kappa)$,
\begin{equation}
\C_{\alpha\beta}(\kappa) \D^{\beta\gamma}(\kappa)= \delta_\alpha^\gamma.
\label{3.10}\end{equation}
Under these assumptions, 
the constrained manifold 
\begin{equation}
 \cK \times A \times \cK := \{ (\kappa^F, g, \kappa^I)\}
\label{3.11}\end{equation}
is equipped with an induced Poisson bracket $\{\ ,\ \}^*$
given by the application of Dirac's well-known formula.
For functions $F_1$ and $F_2$ on  $\cK \times A \times \cK$, we have
\begin{equation}
\{ F_1, F_2\}^*=
 \{ \tilde F_1, \tilde F_2\} - 
\{ \tilde F_1, \lambda^I_\alpha\} \D^{\alpha\beta}(\kappa^I) 
\{ \lambda^I_\beta, \tilde F_2\}
+\{ \tilde F_1, \lambda^F_\alpha\} \D^{\alpha\beta}(\kappa^F) 
\{ \lambda^F_\beta, \tilde F_2\},
\label{3.12}\end{equation}
where the $\tilde F_i$ are arbitrary extensions
of the $F_i$ to a neighbourhood of the constrained manifold
in $\cL \times A \times \cL$ 
and the function on the right hand side is restricted
to $\cK \times A\times \cK$ after the evaluation of the Poisson brackets.
Convenient extensions are provided by requiring the $\tilde F_i$ to
be independent of $\mu^I$ and $\mu^F$ defined by (\ref{3.5}).
Proceeding in this manner, we easily find the following  
Dirac brackets:
\begin{eqnarray}
&&
\{ g_1, g_2\}^* = g_1 g_2 r^*(\kappa^I) - 
r^*(\kappa^F) g_1 g_2 
\nonumber\\
&& \{ g, \kappa^I_i\}^* = g K_i
\nonumber\\
&& \{ g, \kappa^F_i\}^* = K_i g
\nonumber\\
&& \{ \kappa^I_i, \kappa^I_j\}^* =-f_{ij}^{\phantom{ij}k}
\kappa^I_k
\nonumber\\
&& \{ \kappa^F_i, \lambda^F_j\}^* =f_{ij}^{\phantom{ij}k}
\kappa^F_k
\nonumber\\
&&\{ \kappa^I_i, \kappa^F_j\}^* =0,
\label{3.13}\end{eqnarray}
where 
\begin{equation}
r^*(\kappa):= r(\kappa) + \D^{\alpha\beta}(\kappa) M_\alpha \otimes M_\beta
\qquad
\forall \kappa\in \cK.
\label{3.14}\end{equation}
Notice that the Dirac brackets that involve 
the components of $\kappa^I$ or $\kappa^F$ are `the same' as the corresponding
original Poisson brackets.
This is guaranteed by (\ref{3.12}) upon using (\ref{3.1}),
which explains why this assumption was made.
For later reference,  denote the restriction of $r: \cL \rightarrow \A\otimes \A$
to $\cK$ by $\tilde r$ and introduce the map 
$\D: \cK \rightarrow \A\otimes \A$ in correspondence with the second term in (\ref{3.14}).
In this notation,  
\begin{equation}
r^* = \tilde r + \D. 
\label{3.15}\end{equation}
It is obvious that the symmetric part of $r^*$ equals the symmetric part of $r$,
which is an $A$-invariant constant. 
The $\K$-equivariance of the map $r^*: \cK \rightarrow \A\otimes \A$ 
is guaranteed since the Dirac bracket satisfies the Jacobi identity, and
one can also directly check this equivariance property:
\begin{equation}
[ K_i^1 + K_i^2, r^*(\kappa) ]= f_{ji}^{\phantom{ji}k}\kappa_k 
\frac{\partial r^*(\kappa)}{ \partial \kappa_j}. 
\label{3.16}\end{equation} 
For the same reason, it follows that 
\begin{equation}
CDYB(r^*):= 
[r^*_{12}, r^*_{13}] + K^1_i \frac{\partial r^*_{23}}{\partial \kappa_i} +\hbox{cycl. perm.}
\label{3.17}\end{equation}
defines an $A$-invariant constant element of $\A\otimes \A\otimes \A$.
One may expect this constant to be the same as the constant 
given by $CDYB(r)$, which is determined by the formula (\ref{1.3}).
Indeed, we can verify the following statement.
 
\medskip
\noindent 
{\bf Proposition 1.} {\em Consider an $\L$-equivariant map 
$r: \check \L^*\rightarrow \A\otimes \A$
and suppose that equations (\ref{3.1}) and (\ref{3.9}) hold. 
Then one has the equalities 
\begin{equation}
CDYB(\D)=0, 
\qquad\qquad
CDYB(r^*) = CDYB(r),
\label{3.18}\end{equation}
where $r^*: \check\K^* \rightarrow  \A\otimes \A$ and 
$\D: \check\K^*\rightarrow  \M\otimes \M$ are given by (\ref{3.14}) with (\ref{3.10}).}
\medskip

This statement and its interpretation in terms of Dirac reduction 
represent the first main result of the present paper.
The first equality means that under the assumptions in (\ref{3.1}) and (\ref{3.9})
the map $\D: \cK \rightarrow \A\otimes\A$ is an antisymmetric solution
of the CDYBE for the pair $\K\subset \A$.
More precisely, since $\D(\cK)\subset \L \otimes \L$,
this is a solution of the CDYBE for the pair $\K\subset \L$. 
The second equality implies that if $r$ is a classical dynamical $r$-matrix
for the pair $\L\subset \A$ then so is $r^*$ for the pair $\K\subset \A$.
In the special case for which $\K$ is a Cartan subalgebra and $\L$ 
is a reductive subalgebra of a simple Lie algebra, 
the statement of the proposition had been proved in \cite{EV}.
In fact, the proof that we present is analogous to the
proof of Theorem 3.14 in \cite{EV}, but we use only the assumptions in
(\ref{3.1}) and (\ref{3.9}) without any other special features of the Lie algebras 
$\K\subset \L\subset \A$.

In order to verify that $CDYB(\D)=0$, first note that
\begin{equation}
K_i^1 \frac{\partial \D_{23}(\kappa)}{\partial \kappa_i}
=-f_{\gamma\theta}^{\phantom{\gamma\theta}i} 
 \D^{\gamma\alpha}(\kappa) \D^{\theta \beta}(\kappa)
K_i\otimes M_\alpha\otimes M_\beta,
\label{3.19}\end{equation}
which follows by computing the derivatives on account of (\ref{3.10}) and (\ref{3.8}).
By using this, we find that
\begin{equation}
CDYB(\D)= Q^{\nu\alpha\beta} M_\nu \otimes M_\alpha\otimes M_\beta
\label{3.20}\end{equation}
with 
\begin{equation}
Q^{\nu\alpha\beta}
=f_{\gamma\theta}^{\phantom{\gamma\theta}\nu}\D^{\gamma\alpha}\D^{\theta\beta}
+f_{\gamma\theta}^{\phantom{\gamma\theta}\beta}\D^{\gamma\nu}\D^{\theta\alpha}
+ f_{\gamma\theta}^{\phantom{\gamma\theta}\alpha}\D^{\gamma\beta}\D^{\theta\nu}.
\label{3.21}\end{equation}
Multiplying by invertible matrices, we then obtain
\begin{equation}
Q^{\nu\alpha\beta}\C_{\alpha\xi}\C_{\beta\eta} =
f_{\xi\eta}^{\phantom{\xi\eta}\nu} -  \D^{\gamma\nu}
(f_{\eta\gamma}^{\phantom{\eta\gamma}a}f_{a\xi}^{\phantom{a\xi}i}+
f_{\gamma\xi}^{\phantom{\gamma\xi}a}f_{a\eta}^{\phantom{a\eta}i})\kappa_i
=f_{\xi\eta}^{\phantom{\xi\eta}\nu} +  \D^{\gamma\nu} 
f_{\xi\eta}^{\phantom{\xi\eta}a}f_{a\gamma}^{\phantom{a\gamma}i}\kappa_i=0.
\label{3.22}\end{equation}
For the first equality, we used that
$f_{\eta\gamma}^{\phantom{\eta\gamma}a}f_{a\xi}^{\phantom{a\xi}i}
=f_{\eta\gamma}^{\phantom{\eta\gamma}\alpha}f_{\alpha\xi}^{\phantom{\alpha\xi}i}$,
where the indices $a$ and $\alpha$ run over the bases of $\L$ and $\M$,
respectively, and the equality holds because of (\ref{3.1}).
The second equality is valid on account of the Jacobi identity for $\L$,
while the third equality is implied by the definitions of $\C$ and $\D$.
Since we have shown that $Q^{\nu\alpha\beta}=0$, $CDYB(\D)=0$ follows by (\ref{3.20}). 

We start the proof of the second equality in (\ref{3.18}) by remarking that
\begin{equation}
[M^2_\alpha + M^3_\alpha, r_{23}(\kappa)]= \C_{\alpha\beta}(\kappa)
\frac{\partial r_{23}}{\partial \lambda_\beta}(\kappa).  
\label{3.23}\end{equation}
This follows from (\ref{2.6}) upon imposing the constraint 
$\lambda = \kappa\in \cK$.
This equality then implies that
\begin{equation}
M^1_\alpha\frac{\partial r_{23}}{\partial \lambda_\alpha}(\kappa)
= [ \D_{12}(\kappa) + \D_{23}(\kappa), r_{23}(\kappa)].
\label{3.24}\end{equation}
By using this and $CDYB(\D)=0$, it is easy to obtain from (3.15) that 
\begin{equation}
[r_{12} , r_{13}](\kappa) + L_a^1 
\frac{\partial r_{23}}{\partial \lambda_a}(\kappa)
+\hbox{cycl. perm.}= [r^*_{12}(\kappa), r^*_{23}(\kappa)] +
K_i^1 
\frac{\partial r^*_{23}}{\partial \kappa_i}(\kappa)
+\hbox{cycl. perm.},
\label{3.25}\end{equation}
whereby the proof is complete.

   For any constant, nonzero, $A$-invariant element $\varphi \in \A\wedge\A\wedge \A$,
a modified version of the CDYBE may be defined 
by replacing  the zero on the right hand side of (\ref{1.1}) with $\varphi$.
It is clear from Proposition 1 that the Dirac reduction maps not only
the solutions of the CDYBE but also 
the solutions of this modified CDYBE with respect to $\L\subset \A$
to those with respect to $\K\subset \A$, for any fixed invariant $\varphi$.

\section{Examples on self-dual Lie algebras}
\setcounter{equation}{0}

The examples that we describe next are obtained in the situation
for which the Lie algebra $\A$ admits
a nondegenerate, invariant
symmetric bilinear form\footnote{Such Lie algebras,
which include e.g. the reductive Lie algebras and the Drinfeld doubles 
of the Lie bialgebras,  are called self-dual in this paper.
For their structure, one may consult \cite{Figu}.}
 $\langle\ ,\ \rangle$ that remains nondegenerate upon restriction
to the subalgebras $\K$ and $\L$ in the chain
$\K\subset \L \subset \A$.
The bilinear form induces the identifications $\A^*=\A$, $\L^*=\L$, $\K^*=\K$
and allows one to associate with any element of $\A\otimes \A$ a
linear operator on $\A$; 
the operator associated with $X\otimes Y$ sends $Z$
to $\langle Y,Z\rangle X$ for any $X,Y,Z\in \A$.
   The assumption in (\ref{3.1}) is now guaranteed if we let 
\begin{equation}
\M:=\K_\L^\perp := \{ \lambda \in \L\,\vert\, 
\langle\lambda, \kappa\rangle =0 \quad \forall \kappa\in \K\}.  
\label{4.1}\end{equation}
Let us now suppose that the invertibility assumption (\ref{3.9}) holds and
denote the $\mathrm{End}(\A)$ valued functions 
associated with $r$ and $r^*$ by $\rho$ and $\rho^*$,
respectively.
Then formula (\ref{3.15}) can be rewritten in the form
\begin{equation}
\rho^*(\kappa)(X)=\left\{
\begin{array}{cc} 
\rho(\kappa)(X) &\mbox{if\,\, $X\in (\K+ \L^\perp)$}\\
\rho(\kappa)(X)+
\Bigl( \left(\mathrm{ad}\, 
\kappa\right)\vert_{\K^\perp_\L} \Bigr)^{-1}(X) &\mbox{if\,\, $X\in \K^\perp_\L$}.
\end{array}\right. 
\label{4.2}\end{equation}
The domain $\check \K$ 
consists of those elements $\kappa\in \check \L \cap \K$ for
which the restriction of the operator $\mathrm{ad}\, \kappa$ to
$\K^\perp_\L$ is invertible.
To obtain concrete examples, we have to start with a dynamical $r$-matrix
$\rho: \check \L\rightarrow \mathrm{End}(\A)$ 
and have to ensure that $\check \K$ is nonempty.

If we start with the trivial (zero) $r$-matrix, 
then {\em (\ref{4.2}) with $\rho=0$ provides an antisymmetric solution
of the CDYBE whenever $\check \K\subset \K$ is nonempty.}
Although this remark appears quite trivial,  many  
antisymmetric solutions of the CDYBE can be understood as its special cases.
For example,  Theorem 3.2 of  \cite{EV} implies that
if one takes $\K$ to be a Cartan subalgebra of a simple Lie algebra $\A$ and let $\L$ vary, 
then one can recover from (\ref{4.2}) essentially (i.e. up to some
obvious gauge transformations) all 
antisymmetric solutions of the CDYBE for the pair $\K\subset \A$. 

Somewhat more interestingly, 
we may also take as our starting point  a `canonical' dynamical $r$-matrix 
that is available 
in the case $\L=\A$ for {\em any} self-dual Lie algebra $\A$.
This $r$-matrix is defined by using the holomorphic complex function
\begin{equation}
f(z) := \frac{1}{2}\coth\frac{z}{2} - \frac{1}{z}.
\label{4.3}\end{equation}
It was found  
in \cite{EV,AM,BFP} (see also \cite{ES+,FPlong}) 
that the $r$-matrix associated 
with the linear operator
\begin{equation}
\rho_\pm (\lambda) = f(\ad \lambda)  \pm \frac{1}{2} I,
\qquad
\lambda \in \check \A
\label{4.4}\end{equation}
solves the CDYBE (\ref{1.1}) for $\L=\A$.
In the standard manner (see e.g.~\cite{Dun}, Chapter VII), 
the holomorphic function $f$ can be applied 
to the operator $\ad \lambda$ with the aid of the formula
\begin{equation}
f(\ad \lambda) := \frac{1}{2\pi i} \oint_\Gamma dz f(z) (z I - \ad \lambda)^{-1},
\label{4.5}\end{equation}
where $\Gamma$ is a contour that encircles each eigenvalue of $\ad \lambda$.
This expression is well-defined and is independent of the contour $\Gamma$ if $f$
is holomorphic in a neighbourhood of the spectrum $\sigma_\lambda$ of $\ad \lambda$.
Thus the domain $\check \A$ in this case is naturally specified as
\begin{equation}
\check \A = \{ \lambda\in \A\,\vert\, 2\pi i n \notin \sigma_\lambda 
\quad \forall n\in \bZ^*\,\},
\qquad\quad
 (\bZ^*=\bZ\setminus \{0\}).
\label{4.6}\end{equation}
The definition of $f(\ad \lambda)$ in (\ref{4.5}) is equivalent 
to the alternative definition by means of the Taylor series expansion 
of $f(z)$ around $z=0$, which is applicable if $\sigma_\lambda$ lies inside
the disc on which that series converges \cite{Dun}.

In terms of the corresponding linear operators $\rho_\pm^*$,
the reduction of the $r$-matrix (\ref{4.4}) 
to a self-dual subalgebra $\K\subset \A$ 
(on which $\langle\ ,\ \rangle$ remains nondegenerate)
can now be written as follows:
\begin{equation}
\rho_\pm^*(\kappa)=\left\{
\begin{array}{cc} 
f(\ad \kappa) \pm \frac{1}{2} I &\mbox{on $\K$}\\
\frac{1}{2}\coth\left(\frac{1}{2}\ad \kappa\right) 
\pm \frac{1}{2} I &\mbox{on $\K^\perp_\A$}.
\end{array}\right. 
\label{4.7}\end{equation}
In this equation   
$\frac{1}{2}\coth\left(\frac{1}{2}\ad \kappa\right)= f(\ad \kappa) + 
\left(\ad\kappa\right)^{-1}$ on $\K^\perp_\A$ with $f$ in (\ref{4.3}).
Hence {\em the operators $\rho_\pm^*(\kappa)$ are well-defined
on $\A$ if and only if the spectrum of $\ad \kappa$, acting on $\A$,  
does not intersect
$2\pi i \bZ^*$, and    
$\left(\ad \kappa\right)\vert_{\K^\perp_\A}$ is invertible.
These are the conditions that the domain 
$\check \K\subset \K$ must satisfy.}
The invertibility requirement on 
$\left(\ad \kappa\right)\vert_{\K^\perp_\A}$
means that one must restrict the original domain $\check \A$ 
for the reduction to be performable.
Since $\check \K$ does not contain the zero element,
its nonemptyness is a nontrivial  
condition on the subalgebra $\K\subset \A$. 
To state the result in the alternative tensorial terms, 
if the above conditions are satisfied, then solutions
of the CDYBE for $\K\subset \A$ are provided by the 
functions $r_\pm^*: \check \K \rightarrow \A\otimes \A$ given by
\begin{equation}
r_\pm^*(\kappa)= \pm \frac{1}{2}\hat I +  
\langle K^i, f(\ad \kappa) K^j\rangle K_i \otimes K_j
+ \langle M^\alpha,  
\frac{1}{2}\coth(\frac{\ad \kappa}{2}) 
M^\beta \rangle M_\alpha\otimes M_\beta.
\label{4.8}\end{equation}
Here $K_i, K^j$ and $M_\alpha, M^\beta$ denote dual bases 
of $\K$ and $\K^\perp_\A$, respectively, 
$\langle K_i, K^j\rangle =\delta_{i}^j$ 
and $\langle M_\alpha, M^\beta\rangle =
\delta_\alpha^\beta$,
and $\hat I:=  K_i \otimes K^i +  M_\alpha \otimes M^\alpha$.

As we shall see below, 
many known solutions of the CDYBE can be recovered as
special cases of (\ref{4.7}), (\ref{4.8}).
The class of $r$-matrices given by these equations 
apparently has not been displayed before in this general form;
its full set of special cases is still to be uncovered.
It can be checked 
independently of the Dirac reduction argument, too, that 
(\ref{4.7}) provides a solution 
of the CDYBE whenever one has a decomposition 
$\A= \K + \K_\A^\perp$ such that the
above formula yields a well-defined linear operator on $\A$.  
This is important in view of interesting new examples for which $\A$ 
is infinite dimensional with a finite dimensional $\K$, see Section 5.
 
Several solutions of the CDYBE 
can be obtained from (\ref{4.8}) 
by taking $\K:= \A_0$ with respect to an integral gradation 
$\A=\oplus_{n\in {\bZ}}\A_n$ of $\A$ for which   
$\K_\A^\perp = \oplus_{n\in {\bZ}^*}\A_n$.
One must  choose $\A$ and the gradation
in such a way that a nonempty domain $\check \K\subset \K$ exists
on which the components of $r_\pm^*$ are smooth functions.
Notice that this is automatically guaranteed if 
$\A$ is a complex simple Lie algebra.
Indeed, in this case the regular semisimple elements form
a dense open submanifold in $\A_0$ for any integral gradation, and thus 
$\ad \kappa$ is invertible on
$\K^\perp_\A=\oplus_{n\in {\bZ}^*} \A_n$ 
if $\kappa$ belongs to a small  ball 
in $\K$ around such a regular element.
An alternative description of precisely these examples 
is provided by Theorem 3.14 in \cite{EV}.
We note in passing that in this case the second class constraints of 
the Dirac reduction can be naturally separated into first class 
constraints and gauge fixing conditions, simply by 
decomposing $\K_\A^\perp$
into positively and negatively graded subspaces.

To recover the basic trigonometric dynamical $r$-matrix from the above 
mentioned examples, let us now take  
$\A$ to be a finite dimensional complex simple Lie algebra 
equipped with the  principal gradation and  identify $\K$ with  
the Cartan subalgebra given by the grade zero elements. 
Then the $M_\alpha$ in (\ref{4.8}) 
can be taken to be the root vectors 
associated with the set of roots  $\Phi$ with respect to $\K\subset \A$,
and (\ref{4.8}) yields
\begin{equation}
r_\pm^*(\kappa)= \pm \frac{1}{2}\hat I 
+ \sum_{\alpha\in \Phi}  
\frac{\vert \alpha\vert^2}{4}\coth(\frac{\alpha(\kappa)}{2}) 
M_\alpha\otimes M_{-\alpha}.
\label{4.9}\end{equation}   
We here used that $[\kappa, M_\alpha]=\alpha(\kappa) M_\alpha$, 
$\langle M_\alpha, M_{-\alpha}\rangle = \frac{2}{\vert \alpha \vert^2}$
and that $f(\ad\kappa)K^j$ in (\ref{4.8}) now vanishes since $\K$ is Abelian.
This solution of the CDYBE (\ref{1.1}) first appeared in 
studies of the WZNW and conformal
Toda field theories \cite{BDF}. 
It has been proved in \cite{EV} that if
the dynamical variable belongs to a Cartan subalgebra of a simple Lie algebra  
then all solutions of  (\ref{1.1}) 
can be obtained from (\ref{4.9}) by shifts of
the argument $\kappa$ by a constant and simple limiting procedures.  
Note in passing that the analogous reduction with $r=0$ as starting point
leads to the rational $r$-matrix \cite{EV} 
$r^*_{0}(\kappa)=\sum_{\alpha\in \Phi} 
\frac{\vert \alpha\vert^2}{2\alpha(\kappa)} M_\alpha\otimes M_{-\alpha}$.

It was found in \cite{EV} that the natural generalization 
of  (\ref{4.9})
defines a dynamical $r$-matrix also for an affine Kac-Moody Lie algebra $\A$.
To obtain this generalization, 
one uses the principal gradation of $\A$ for which $\K:=\A_0$ 
is the Cartan subalgebra, and correspondingly extends the summation in  
(\ref{4.9})  over the roots of $\A$.
Motivated by this result, in Section 5
we display a large family of dynamical $r$-matrices 
on affine Lie algebras based on {\em arbitrary} finite 
dimensional self-dual Lie algebras.

Before turning to infinite dimensional Lie algebras, 
we wish to show that our general formula (\ref{4.7})  
contains new examples for finite dimensional self-dual Lie algebras, too. 
To illustrate this,  we now take $\A$ to be the well-known \cite{Sfetsos}
self-dual extension of the complex Euclidean Lie algebra $e(d)$.
We denote the generators of $e(d)$ ($d\geq 2$) by $P_i$ and $J_{ij}$, where 
$i,j=1,\ldots,n$ and the relation $J_{ji}= - J_{ij}$ is understood. 
The $J_{ij}$ span the orthogonal Lie algebra $o(d)\subset e(d)$, and we 
let $T_{ij}$ $( T_{ji}=-T_{ij})$ denote the generators 
of the dual space of $o(d)$. 
By definition, $\A={\rm span}\{ P_i, J_{ij}, T_{ij}\}$ has 
the commutation relations 
\begin{eqnarray}
&& [J_{ij}, J_{kl}]= \delta_{jk} J_{il} + \delta_{il} J_{jk}
-\delta_{jl} J_{ik}-\delta_{ik} J_{jl}\nonumber\\
&& [J_{ij}, T_{kl}]= \delta_{jk} T_{il} + \delta_{il} T_{jk}
-\delta_{jl} T_{ik}-\delta_{ik} T_{jl}\nonumber\\
&& [T_{ij}, T_{kl}]= [T_{ij}, P_k]=0\nonumber\\
&& [ J_{ij}, P_k] = \delta_{jk} P_i - \delta_{ik} P_j \nonumber\\
&& [P_k, P_l] = T_{kl}. 
\label{4.10}\end{eqnarray}
There is a one-parameter family of invariant scalar products on $\A$, 
which in terms of our redundant set of generators is given by 
\begin{eqnarray}
&&\langle J_{ij}, J_{kl} \rangle = 
p (\delta_{jk} \delta_{il} - \delta_{ik} \delta_{jl})\nonumber\\
&&\langle J_{ij}, T_{kl} \rangle = 
\delta_{jk} \delta_{il} - \delta_{ik} \delta_{jl}\nonumber\\
&&\langle J_{ij}, P_k \rangle = \langle T_{ij}, P_k \rangle=0\nonumber\\
&& \langle P_{i}, P_j \rangle = \delta_{ij},
\label{4.11}\end{eqnarray}
where $p$ is an arbitrary constant.
It is clear that $\K:={\rm span}\{ J_{ij}, T_{ij}\}$ 
is a self-dual subalgebra
and $\K_\A^\perp= {\rm span}\{ P_i \}$. 
By writing 
$\kappa \in \K$ as $\kappa = x^{ij} J_{ij} + y^{ij} T_{ij}$ and 
$P\in \K_\A^\perp$ as $P= z^i P_i$, where summation is understood and
the components $x^{ij}, y^{ij}$ are antisymmetric in $ij$, we see that  
$[\kappa, P] = 2\sum_{i,k} x^{ik} z^k  P_i$. 
Since the determinant of an antisymmetric matrix of odd size is zero,
$\ad \kappa$ is never invertible on $\K^\perp_\A$ if $d$ is odd,
and hence we do not obtain a nonempty domain $\check \K$ in this case.
However, 
if $d$ is even, then one may check that for
$
\kappa_0 := J_{12} + J_{34} +\cdots + J_{d-1, d}$ the operator 
$(\ad \kappa_0)\vert_{\K_\A^\perp}$ is invertible.
It follows that for a small but nonzero constant 
$q$ the element $q \kappa_0 \in \K$ satisfies the spectral conditions 
described below equation (\ref{4.7}).
This implies that $\check \K \subset \K$ is a {\em nonempty open domain 
for any even $d$}, and (\ref{4.7}) provides us with new 
dynamical $r$-matrices in this case. 

We note that for $d=2$  (4.10) defines the
central extension of $e(2)$ that has interesting physical applications.
In this case $\K$ is a two-dimensional Abelian Lie algebra.
Further examples for which $\K$ is two-dimensional and Abelian
can be obtained by taking $\A$ to be the oscillator Lie algebra
generated by $a_i$, $a_i^\dagger$ $(i=1,\ldots, n)$, 
the central element $\hat c$ and the number operator $\hat N$.
With respect to the usual scalar product \cite{Moroz},
$\K= {\rm span}\{ \hat N, \hat c\}$ is a self-dual 
subalgebra and $\check \K$ 
is easily seen to be nonempty.  
In these cases, it should not be too difficult to quantize the above 
constructed dynamical $r$-matrices. 

\section{Generalizations to affine Lie algebras}
\setcounter{equation}{0}

In this section we describe generalizations of the 
$r$-matrices that appear in (4.7) for situations in which 
the dynamical variable lies in a finite dimensional
subalgebra of an  infinite dimensional Lie algebra $\A$. 
In fact, we shall take $\A$ to be an
`affine Lie algebra' obtained by central extension and 
inclusion of the derivation from a twisted
loop algebra built on a finite dimensional self-dual Lie algebra $\G$,
and let the dynamical variable lie in the grade zero part of $\A$.

We start with a preliminary remark that will be used below.
Let $\A$ be a (possibly infinite dimensional) 
self-dual Lie algebra with
scalar product $\langle\ ,\ \rangle$.
Consider a decomposition 
\begin{equation}
\A=\K + \K^\perp,
\qquad
\K\cap \K^\perp=\{0\},
\label{5.1}\end{equation}
where $\K\subset \A$ is a finite dimensional self-dual Lie subalgebra.
Let now denote by $R: \check \K \rightarrow \mathrm{End}(\A)$
the operator valued function corresponding to a 
function\footnote{ 
If $\A$ is infinite dimensional, then 
$\A\otimes \A$ denotes a certain completion 
of the algebraic tensor product, 
which is such that the 
corresponding linear operators are well-defined on $\A$.}  
$r: \check \K\rightarrow \A\otimes \A$,
where $\check \K$ is some open subset of $\K$.
By assuming the existence of the directional derivative
\begin{equation}
(\nabla_T R)(\kappa):= \frac{d}{dt} R(\kappa + t T) \vert_{t=0}
\qquad
\forall T\in \K,\quad \kappa\in \check \K,
\label{5.2}\end{equation}
let us define 
\begin{equation}
\langle X, (\nabla R)(\kappa) Y\rangle  := 
\sum_i K^i \langle X, (\nabla_{K_i} R)(\kappa) Y \rangle,
\qquad \forall X, Y\in \A,
\label{5.3}\end{equation}
where $\{ K_i\}$ and $\{ K^i\}$ are dual bases of $\K$,
$\langle K_i, K^j\rangle = \delta_i^j$.
Denote by $\hat f\in \A\otimes\A\otimes \A$ the 
(antisymmetric) invariant
element 
associated with the Lie bracket of $\A$, and $\hat I \in \A\otimes \A$
the (symmetric) invariant element associated with the unit operator on $\A$.
(If $T_\alpha$ and $T^\alpha$ are dual bases of 
$\A$ and $[T_\alpha, T_\beta] = 
f_{\alpha\beta}^\gamma T_\gamma$, then
$\hat f= f_{\alpha\beta}^\gamma T^\alpha \otimes T^\beta \otimes T_\gamma$ 
and $\hat I= T_\alpha \otimes T^\alpha$.)
We have the following lemma.

\medskip
\noindent
{\bf Lemma.}
{\em Let us consider an antisymmetric r-matrix 
$r: \check \K\rightarrow \A\wedge \A$ and the associated operator
$R: \check \K \rightarrow \mathrm{End}(\A)$. 
Then the equation 
\begin{equation}
CDYB(r)=- C^2 \hat f,
\label{5.4}\end{equation}
where $C$ is some complex constant, 
is equivalent to
\begin{eqnarray}
&&[ R X, R Y] -R( [X, R Y]+ [RX,Y]) 
+\langle X, (\nabla R) Y\rangle + (\nabla_{Y_\K} R) X -
(\nabla_{X_\K} R) Y \nonumber\\
&&\qquad\qquad \qquad = -C^2 [X,Y], \qquad \forall X, Y\in \A.
\label{5.5}\end{eqnarray}
}

The statement of the lemma is straightforward to verify.
Note that in (\ref{5.5}) we use the decomposition 
$X= X_\K + X_{\K^\perp}$
with 
$X_\K\in \K$, $X_{\K^\perp}\in \K^\perp$
and similarly for $Y$.
The variable $\kappa\in \check \K$ 
had been omitted for brevity; $RX$ stands for 
the action of $R(\kappa)$ on $X\in \A$ and so on.
It is often more convenient to verify (\ref{5.5}) case by case 
for the different choices of $X$ and $Y$,
than to inspect all components of the threefold 
tensor product in (\ref{5.4}).
It is well-known that (\ref{5.4}) is also equivalent to
$CDYB(r\pm C \hat I)=0$. 

Let now $\G$ be a finite dimensional complex, self-dual Lie algebra
with the invariant `scalar product' denoted as 
$B(\xi,\eta)$ for any $\xi, \eta\in \G$.
Let us suppose that $\mu$ is an automorphism of $\G$ 
of order $N\in {\bN}$, $\mu^N=\mathrm{id}$, 
that has nonzero fix points and satisfies $B(\mu(\xi), \mu(\eta))= B(\xi,\eta)$.
(The last two properties of $\mu$ are automatic if $\mu=\mathrm{id}$ or $\G$ is simple,
which are included as special cases.)
Then $\G$ can be decomposed as a direct sum of 
the eigensubspaces of $\mu$ as
\begin{equation}
\G= \oplus_{a\in \E_\mu} \G_a,
\qquad
\E_\mu\subset \{ 0, 1, \ldots, (N-1)\,\},
\label{5.6}\end{equation}
\begin{equation}
\G_a:= \{ \xi\in \G\,\vert\, \mu(\xi)= 
\exp(\frac{ia2\pi}{N}) \xi\,\} \neq \{ 0\}.
\label{5.7}\end{equation}
Note that $\G_a$ is perpendicular to $\G_b$ with respect to the bilinear form $B$
unless $a+b=N$ or $a=b=0$, which implies that if a nonzero $a$
belongs to the index set $\E_\mu$ then so does $(N-a)$, 
and $\G_0\neq \{ 0\}$ is a self-dual subalgebra of $\G$.
The {\em twisted loop algebra} $\ell(\G,\mu)$ is by definition 
the subalgebra of $\G\otimes {\bC}[\lambda, \lambda^{-1}]$
generated by elements of the form
\begin{equation}
\xi^{n_a}:= \xi \otimes \lambda^{n_a}
\quad\hbox{with}\quad
\xi\in \G_a,
\quad
n_a = a + m_{a} N,
\,\,\, m_a\in {\bZ}.
\label{5.8}\end{equation}
The `affine Lie algebra' $\A(\G,\mu)$ is given by
\begin{equation}
\A(\G,\mu):= \ell(\G, \mu) 
\oplus {\bC} d \oplus {\bC} \hat c 
\label{5.9}\end{equation}
with the Lie bracket of its generators defined as 
\begin{equation}
[\xi^{n_a}, \eta^{p_b}] = [\xi,\eta]^{n_a + p_b}
+ n_a \delta_{n_a,- p_b} B(\xi,\eta)\hat c,
\quad
\forall \xi\in \G_a,\,\,\,\eta\in \G_b,
\label{5.10}\end{equation}
\begin{equation}
[ d, \xi^{n_a}]= n_a \xi^{n_a},
\quad
[\hat c, d] = [\hat c, \xi^{n_a}]=0. 
\label{5.11}\end{equation}
A nondegenerate scalar product $\langle\ ,\ \rangle$
can be defined on $\A(\G,\mu)$ by setting 
\begin{equation}
\langle \xi^{n_a}, \eta^{p_b}\rangle  = 
\delta_{n_a, -p_b} B(\xi,\eta),
\quad
\langle \hat c, d \rangle =1,
\quad
\langle d, \xi^{n_a}\rangle = \langle \hat c, \xi^{n_a}\rangle =0.
\label{5.12}\end{equation}
Notice that $\A(\G,\mu)$ is graded by the eigenvalues
of $\ad d$, 
\begin{equation} 
\A(\G,\mu)= \oplus_{n\in (\E_\mu + N {\bZ})} \A(\G,\mu)_n,
\label{5.13}\end{equation}
whereby we obtain a decomposition of the type (\ref{5.1}) with
\begin{equation}
\K:= \A(\G,\mu)_0 = \G_0 \oplus {\bC} d \oplus {\bC}\hat c,
\qquad
\K^\perp = \oplus_{n\in (\E_\mu + N {\bZ})\setminus \{ 0\} } \A(\G,\mu)_n.
\label{5.14}\end{equation}
We regard $\G_0$ as a subspace of $\A(\G,\mu)$ by
identifying $\xi\in \G_0$ with $\xi\otimes \lambda^0\in \A(\G,\mu)$;
and now we set $\A:= \A(\G,\mu)$ for brevity.

To  describe the dynamical $r$-matrix
of our interest,  $R: \check \K \rightarrow \mathrm{End}(\A)$,
we parametrize the general element $\kappa\in \K=\A_0$ as
\begin{equation}
\kappa= \omega + k d + l \hat c,
\qquad
\omega\in \G_0,\quad k,l\in {\bC}.
\label{5.15}\end{equation}
Let $f$ and $F$ be the following complex analytic functions:
\begin{equation}
f: z \mapsto \frac{1}{2}\coth \frac{z}{2}- \frac{1}{z},
\qquad
F: z \mapsto \frac{1}{2}\coth \frac{z}{2}.
\label{5.16}\end{equation}
By definition,  $R(\kappa)$ $(\kappa \in \check \K)$ 
is given by the collection of the
 finite dimensional linear operators 
\begin{equation}
R(\kappa)\vert_{\A_0} := f((\ad \kappa)_0),
\quad
R(\kappa)\vert_{\A_n} := F((\ad \kappa)_n) \quad 
\forall n\in\ (\E_\mu + N {\bZ})\setminus \{0\},
\label{5.17}\end{equation}
where 
$(\ad \kappa)_n:= \ad \kappa\vert_{\A_n}$ 
$\forall n\in\ (\E_\mu + N {\bZ})$.
These finite dimensional operators are given analogously to (\ref{4.5}).
Therefore, for them  to be well-defined, the spectrum of $\ad \kappa$ 
on $\A_n$ must not contain any pole of the respective functions 
$f$ (for $n=0$) and $F$ (for $n\neq 0$).
This condition could be spelled out  explicitly by
using that for $\xi \in \G_a$ and $n_a=(a + m N)$ with $m\in \bZ$
$(\ad \kappa ) \xi^{n_a} = \left((k n_a + \ad \omega) \xi\right)^{n_a}$.
This relation translates the condition on the spectrum of $\ad \kappa$ into
a condition on the spectrum of $\ad \omega$ on the $\G_a$.
It is not difficult to see from this that 
$R: \check \K \rightarrow  \mathrm{End}(\A)$ 
is indeed well-defined on a  domain of the form
\begin{equation}
\check \K= \{ \,\kappa = \omega + k d + l \hat c\,\vert\,
l\in {\bC}, \,\,
k \in ({\bC}\setminus {\bR}i),
\,\,\,  \omega \in \B_k \},
\label{5.18}\end{equation}
where $\B_k\subset \G_0$ is an open subset depending on $k$ for 
which the above conditions hold (for a more explicit description, 
see \cite{FPlong}). 
The corresponding map $r: \check \K \rightarrow \A\otimes \A$ 
is antisymmetric and is $\K$-equivariant.  Its interest is due to 
the following statement.

\medskip
\noindent
{\bf Proposition 2.} {\em
The dynamical $r$-matrix $R: \check \K \rightarrow \mathrm{End}(\A)$ 
defined by equations (\ref{5.17}) with (\ref{5.16}) on a domain of the form
in (\ref{5.18}) satisfies the operator version (\ref{5.5}) 
of the CDYBE with $C=\frac{1}{2}$. }
\medskip

The verification of the proposition  is not difficult, but
it is rather long.  It is presented in \cite{FPlong}.
As an equivalent statement, it follows 
that the $r$-matrices $r^\pm: \check \K \rightarrow \A\otimes \A$ that 
are associated with the operators $R^\pm := R \pm \frac{1}{2} I$  
satisfy the CDYBE (\ref{1.1}). 
Formally, these $r$-matrices can be thought of as special cases of (\ref{4.8}). 
Our point is that they are well-defined in the infinite dimensional
situation considered here.
It may also be checked that these $r$-matrices are $\K$-equivariant,
the condition in terms of $R$ being $(\nabla_{[T, \kappa]  } R)(\kappa) = 
[\ad T, R(\kappa)]$, $\forall T\in \K, \kappa\in \check \K$.

We finish by a remark on a reinterpretation of the above 
$\A\otimes \A$-valued $r$-matrices as spectral parameter dependent $r$-matrices.
It is well-known that spectral parameter dependent 
$\G\otimes \G$-valued $r$-matrices 
 may be obtained by applying evaluation homomorphisms
to $\ell(\G,\mu)\otimes \ell(\G,\mu)$-valued $r$-matrices.
In the context of dynamical $r$-matrices, Etingof and Varchenko \cite{EV}
used this method to recover Felder's elliptic dynamical 
$r$-matrices \cite{Feld} from 
the standard trigonometric dynamical $r$-matrices of the 
affine Lie algebras based on the complex simple Lie algebras. 
In fact, the same procedure can be applied to the more general family 
of dynamical $r$-matrices given by Proposition 2.
The first step is to set $\hat c$ to zero and fix the value of $k$.
Thereby  $r^{\pm}(\kappa)\in \A\otimes \A$ become 
$\ell(\G,\mu)\otimes \ell(\G,\mu)$-valued dynamical $r$-matrices,
$r^{k,\pm}: {\cal B}_k \rightarrow \ell(\G,\mu) \otimes \ell(\G,\mu)$,
which depend parametrically on $k$.
By using the standard evaluation homomorphisms along the lines of \cite{EV},
$r^{k, \pm}$ are then converted into $\G\otimes \G$-valued
spectral parameter dependent dynamical $r$-matrices, $r^{k,\pm}(\omega, z)$.
The final result can be described as follows.
Introduce the functions $\chi_{a}(w,z\vert \tau)$ of the 
complex variables $w$, $z$ by
\begin{equation}
\chi_a(w,z\vert \tau):= \exp\left({\frac{2\pi i a z}{N}}\right)
\left( \frac{1}{2\pi i}
\frac{\theta_1(\frac{w}{2\pi i}+\frac{ a}{N}\tau + z\vert\tau)
\theta'_1(0\vert\tau)}{\theta_1(z\vert \tau) 
\theta_1(\frac{w}{2\pi i}+\frac{ a}{N}\tau \vert\tau)} -\frac{\delta_{a,0}}{w}\right),
\label{5.19}\end{equation}
where $\theta_1$ is the standard 
theta-function\footnote{We have 
$\theta_1(z\vert \tau)=\vartheta_1(\pi z\vert \tau)$ 
with $\vartheta_1$ in \cite{WW}.}, and let
$T_\alpha$, $T^\beta$ denote dual bases of $\G$.
In fact, one obtains the $r$-matrix
$r^{k,+} (\omega, z)= B( T_\alpha, 
{\cal R}(\omega, z\vert \tau) T_\beta) 
T^\alpha \otimes T^\beta$ 
where  ${\cal R}(\omega, z\vert \tau)\in 
\mathrm{End}(\G)$ is defined by 
\begin{equation}
{\cal R}(\omega, z\vert \tau)\vert_{\G_a} := 
\chi_a(\ad \omega, z\vert \tau)\quad\hbox{on}\quad \G_a\quad
\forall a\in \E_\mu,\,\, \omega\in \B_k\subset \G_0.
\label{5.20}\end{equation} 
The relation between the parameters $k$ and $\tau$ reads as
$\tau:= \frac{kN}{2\pi i}$, where we assumed that $\Re (k) <0$.
The derivation of this formula is contained in \cite{FPlong}. 
If $\G$ is a simple Lie algebra and $\mu$ 
is an inner automorphism corresponding to a Coxeter 
element in the Weyl group, 
then the spectral parameter dependent
$r$-matrices given by (\ref{5.20}) are equivalent 
to Felder's elliptic 
dynamical $r$-matrices, as expected upon 
comparison with Section 4.6 in \cite{EV}.
In the general case, the $r$-matrices 
provided by (\ref{5.17}), (\ref{5.20}) appear to be new.

\section{Discussion}

In this paper we pointed out that solutions of 
the CDYBE can be mapped 
to other solutions by  Dirac
reductions of their underlying Poisson-Lie groupoids, if 
the conditions given in (\ref{3.1})
and (\ref{3.9}) are satisfied.
Among the possible applications of Proposition 1, we mentioned the
antisymmetric solutions that are obtained as reductions of the zero 
$r$-matrix\footnote{Note added: We learned after submitting this article
that these dynamical $r$-matrices, 
given by  $\D$ in Proposition 1, have also been found recently in \cite{Xu}
by using a different method.}, 
and the class of $r$-matrices given by (\ref{4.7}) that are 
reductions of the canonical $r$-matrix (\ref{4.4}).
Many of these $r$-matrices are already known, 
but the class defined by (\ref{4.7}) contains new examples, too.
Although our construction works rigorously only in the 
finite dimensional case,
some interesting $r$-matrices that result from it turned out
to be well-defined in certain infinite dimensional situations as well.
In particular, we exhibited a family of dynamical $r$-matrices 
on affine Lie algebras based on arbitrary finite dimensional self-dual Lie
algebras. 
The $r$-matrices provided by Proposition 2 are in the general case new,
and this family includes as special cases those trigonometric 
$r$-matrices of \cite{EV} that  
become Felder's elliptic dynamical $r$-matrices \cite{Feld} 
upon evaluation homomorphisms. 

The `canonical'  $r$-matrices (\ref{4.4})   appear in the description 
of the chiral sectors of the classical WZNW model in association with any 
finite dimensional self-dual Lie algebra \cite{BFP}.
Some of their reductions to self-dual subalgebras have been 
considered in this context in \cite{Dub}.
We also wish to mention the paper \cite{FatLuk},
where the effect of a Dirac reduction 
of the chiral WZNW phase space on constant exchange  
$r$-matrices  has been studied.
The derivation of the trigonometric $r$-matrix (\ref{4.9})
contained in this paper served as one of our original motivations,
but its status  in terms of the geometric 
interpretation of the dynamical $r$-matrices is still to be understood.
Another open question is if the general family of $r$-matrices 
given by Proposition 2 can be used to encode the Poisson brackets of 
generalized versions of the WZNW model. 
As candidates, we have in mind both the
WZNW models formally obtained  by replacing the finite dimensional
WZNW group with an extended affine Lie group \cite{Arat}, which is useful in 
the theory of soliton equations \cite{Fer}, 
and the intriguing quasitriangular WZNW model recently 
introduced by Klimcik \cite{Klim}.
Felder's $r$-matrices are already known to play this role in these models.

Somewhat implicitly (as a special case of Theorem 3.14),
the canonical $r$-matrices (\ref{4.4}) first appeared in \cite{EV}.
In their explicit form,  they were found 
independently in the papers \cite{AM,BFP}.
More precisely, in \cite{EV,AM} the assumption that 
the underlying Lie algebra is reductive was used,
while \cite{BFP} 
provides an indirect approach to these $r$-matrices on 
any self-dual Lie algebra.
A direct proof of the statement 
that (\ref{4.4}) satisfies the CDYBE for any finite dimensional
self-dual Lie algebra is given in \cite{FPlong}.
For a different proof in a generalized case, see \cite{ES+}.
The relationship between the
generalizations of the $r$-matrices (\ref{4.4}) constructed in \cite{ES+}  
and the $r$-matrices given by our Proposition 2 is explained in \cite{FPlong}.

It would be interesting to develop 
the quantization of the $r$-matrices defined by (\ref{4.4}).
If that was found, one could in principle obtain 
the quantizations of the reduced $r$-matrices in (\ref{4.7}) 
by means of appropriate quantum Hamiltonian reductions.
We also wish to clarify  if the canonical $r$-matrices  
make sense in cases for which the dynamical variable $\lambda$ lies
in a `suitable' {\em infinite dimensional} self-dual Lie algebra $\A$. 
Formula (\ref{4.5}) itself is well-defined \cite{Dun}
if $\A$ is a Banach space and $\ad \lambda$ is a bounded operator on it.

We hope to be able to return to the above questions in the future.

\bigskip
\bigskip
\noindent{\bf Acknowledgements.}
L.F. wishes to thank J. Balog for useful discussions and for comments 
on the manuscript. 
This investigation was supported in part by the Hungarian 
Scientific Research Fund (OTKA) under T034170, 
T029802, T030099  and M028418.

\end{document}